\documentclass[11pt,a4paper]{article}

\usepackage[truedimen,margin=32mm]{geometry} 

\usepackage{mathrsfs}
\usepackage{amssymb}
\usepackage{amsmath}
\usepackage{ascmac}
\usepackage{amsthm}
\usepackage[pdftex]{graphicx}
\usepackage[pdftex]{color}
\usepackage{natbib}
\usepackage{setspace}
\usepackage{comment}

\usepackage{color}

\usepackage{titlesec}
\titleformat*{\section}{\large\bfseries}
\titleformat*{\subsection}{\it}

\newtheorem{thm}{Theorem}

\newtheorem{algo}{Algorithm}

\def\ep{{\varepsilon}}

\def\th{{\theta}}

\def\yh{{\widehat{y}}}

\def\thh{{\widehat{\theta}}}

\def\nut{{\widetilde{\nu}}}

\def\st{{\widetilde{s}}}
\def\mt{{\widetilde{m}}}

\def\At{{\widetilde{A}}}
\def\Bt{{\widetilde{B}}}
\def\Mt{{\widetilde{M}}}
\def\Ut{{\widetilde{U}}}
\def\Vt{{\widetilde{V}}}
\def\St{{\widetilde{S}}}

\def\Dc{{\mathcal{D}}}

\title{{\bf Semiparametric Imputation using Latent Sparse Conditional Gaussian Mixtures for Multivariate Mixed Outcomes}}

\date{}

\begin{document}

\maketitle
\doublespacing

\vspace{-1.5cm}
\begin{center}
Shonosuke Sugasawa$^1\footnote{Corresponding author (Email: sugasawa@csis.u-tokyo.ac.jp)}$, Jae Kwang Kim$^2$ and Kosuke Morikawa$^3$
\end{center}

\noindent
$^1$Center for Spatial Information Science, The University of Tokyo\\
$^2$Department of Statistics, Iowa State University\\
$^3$Graduate School of Engineering Science, Osaka University

\vspace{1cm}
\begin{center}
{\bf \large Abstract}
\end{center}

\vspace{-0cm}
This paper proposes a flexible Bayesian approach to multiple imputation using conditional Gaussian mixtures. 
We introduce novel shrinkage priors for covariate-dependent mixing proportions in the mixture models to automatically select the suitable number of components used in the imputation step. 
We develop an efficient sampling algorithm for posterior computation and multiple imputation via Markov Chain Monte Carlo methods. 
The proposed method can be easily extended to the situation where the data contains not only continuous variables but also discrete variables such as binary and count values. 
We also propose approximate Bayesian inference for parameters defined by loss functions based on posterior predictive distributing of missing observations, by extending bootstrap-based Bayesian inference for complete data.
The proposed method is demonstrated through numerical studies using simulated and real data. 

\bigskip\noindent
{\bf Key words}: missing at random; shrinkage prior; sparse mixture

\newpage
\section{Introduction}

Missing data is frequently encountered in statistics. Handling missing data involves some model assumptions about the sample data and the mechanism for creating missingness. The observed likelihood with missing data often requires the computation of complicated integration with respect to missing data. Multiple imputation, however,  predicts the missing values and creates easy-to-use complete data multiple times in a Bayesian way. The estimator in multiple imputation is just the average of estimators computed with each the complete data; hence it is easy to conduct. The large-sample theory for the multiple imputation estimator is established in \citet{wang1998} and \citet{nielsen03}. An alternative advantage of multiple imputation beyond its user-friendliness is its convenience for statistical inference. The Rubin's formula enables us to conduct statistical inference without directly computing the asymptotic variance of the estimator \citep{rubin1987, rubin1996}. 

As for imputation model, instead of making strong parametric model assumptions, it is desirable to use more flexible models so that the resulting analysis can be justified without spending too much time on model diagnostics. Gaussian mixture model is a flexible model assumption widely used in practice \citep{bacharoglou2010}. \cite{kim2014multiple}  proposed Bayesian multiple imputation using the Dirichlet process mixture. 
\cite{sang2020semiparametric} proposed semiparametric fractional imputation using Gaussian mixtures. Recently, \cite{lee2022semiparametric} proposed more flexible model, called  conditional Gaussian mixture model (CGMM), to develop a semiparametric fractional imputation. In the semiparametric imputation, the number of model parameters can also increase with the sample size. However, the uncertainty quantification (i.e. variance estimation) with the semiparametric fractional imputation in \cite{lee2022semiparametric}  is challenging because the parameters are estimated deterministically using the frequentist framework and its uncertainty is not directly captured in the frequentist approach. Also, the computational cost for model selection in \cite{lee2022semiparametric} is based on the BIC criterion, and the computation associated with computing the BIC for many different models can be huge. Furthermore, the GMM or CGMM method cannot handle mixed outcomes since they implicitly assume that the variables are continuous-valued.

In this paper, we first develop a Bayesian version of the CGMM imputation so that the resulting Bayesian inference can be more straightforward.  
The proposed method can automatically tune the number of mixture components by specifying a large number of components and shrinking unnecessary components within MCMC.
It can save computational cost and leads to flexible and efficient multiple imputation. 
We also consider an extension of the proposed method to handle mixed outcomes by using the sparse Gaussian mixtures as the latent variable models. 
Uncertainty in estimating the imputation models can be incorporated into the final posterior distributions of parameters that we are interested in.    
Unlike the famous MICE method of \cite{buuren2011mice}, our proposed method does not suffer from the model incompatibility problem \citep{hughes2014}.  
The proposed method can easily handle mixed outcomes, and the model is clearly compatible.

Another advantage is that the proposed method is computationally attractive. 
The posterior computation can be efficiently carried out via a simple Gibbs sampling algorithm. The proposed method can be used to make posterior inference about the parameters defined through loss-type functions. Furthermore, because the proposed method implements Bayesian imputation, the statistical inference can be easily implemented using Rubin's formula in multiple imputation. However, it requires a condition called ``congeniality" that imposes models used for imputation and analysis being the same \citep{meng1994}. See \citet{xie2017} for a comprehensive discussion on congeniality. 
For general-purpose estimation, variance estimation using Rubin's formula does not necessarily lead to consistent estimation \citep{yang2016}. 
\citet{vonhippel202l} proposed a modified formula to conduct a consistent variance estimation. 
In this paper, instead of using Rubin's formula,  we propose a consistent variance estimator without relying on the congeniality condition by extending the loss-likelihood bootstrap \citep{lyddon2019general} and weighed likelihood bootstrap \citep{newton1994approximate}.
While there are some existing Bayesian approaches to multiple imputation \citep[e.g.][]{kato2020semiparametric, kim2014multiple, murray2016multiple}, they did not consider a valid algorithm for computing posterior distributions of parameter defined through general objective functions.

The paper is organized as follows. In Section 2, the proposed method is presented. In Section 3, an extension to mixed data imputation is described.  In Section 4, we present how to apply the proposed method to different parameters of interest. Results from two limited simulation studies are presented in Section 5. 
A real data application is presented in Section 6. Some concluding remarks are made in Section 7.

\section{Conditional Gaussian mixture model}

\subsection{Setup}
Suppose that $x$ and $y$ are observed in the sample, where $y=(y_1,\ldots,y_p)^T$ is a $p$-dimensional vector of study variables and $x=(x_1,\ldots,x_q)^T$ is a $q$-dimensional vector of auxiliary variables. 
We first assume that all the elements of $y$ are continuous-valued, and a mixed outcome case will be discussed in Section \ref{sec:mixed}.

Let $y_{obs}$ and $y_{mis}$ be the observed and missing part of $y$, respectively, namely, $y=(y_{obs}^T, y_{mis}^T)^T$.
We assume that the missing mechanism is missing at random in the sense of \cite{rubin1976inference}, that is, $f(\delta|x,y)=f(\delta|x,y_{obs})$, where $\delta=(\delta_1,\ldots,\delta_p)^T$ is the response indicator vector for $y$, defined as $\delta_j=1$ if $y_j$ is observed, otherwise $\delta_j=0$.
Then, the imputation model can be carried out via the conditional distribution of $y_{mis}$ given $(x,y_{obs})$ with its density 
$$
f(y_{mis}|x,y_{obs})=\frac{f(y|x)}{\int f(y|x)dy_{mis}}.
$$
If $f(y|x)$ is estimated by a parametric model with parameter $\theta$, the parametric imputation is performed in two steps: estimation of the parameter $\theta$ and generate $y_{mis}$ from the above conditional density. 
\cite{kim2011} proposed parametric fractional imputation using parametric models directly. 
Some flexible approaches for the imputation model are recently proposed \citep[e.g.][]{lee2022semiparametric,sang2020semiparametric} for robust estimation.

\subsection{Sparse Conditional Gaussian Mixtures}\label{sec:SCGM}
We first consider the following conditional Gaussian mixture models for $f(y|x)$ as adopted in \cite{lee2022semiparametric}:
\begin{equation}\label{cond}
f(y|x)=\sum_{g=1}^G\pi_g(x)f_g(y|x;\psi_g),
\end{equation}
where $\pi_g(x)=P(z=g|x)$ and $f_g$ is a Gaussian distribution given $x$ and $z=g$.
We adopt $f_g(y|x;\psi_g)=\phi_p(y; b_g +B_gx, \Sigma_g)$, where $B_g$ is a $(p,q)$-matrix of coefficients and $\Sigma_g$ is a variance-covariance matrix. 
Under the model (\ref{cond}), the conditional probability of $z=g$ is obtained as 
\begin{equation}\label{cond.prob}
P(z=g|x, y_{obs})=\frac{\pi_g(x)f_g(y_{obs}|x;\psi_g)}{\sum_{g'=1}^G\pi_{g'}(x)f_{g'}(y_{obs}|x;\psi_{g'})},
\end{equation}
where 
$$
f_{g}(y_{obs}|x;\psi_{g})=\int f_g(y|x;\psi_g)dy_{mis}
$$
is still a Gaussian density function. 
Then, given $x$ and $y_{obs}$, the missing response $y_{mis}$ can be imputed using 
\begin{equation}\label{impute}
f(y_{mis}|x,y_{obs})=\sum_{g=1}^G\pi_g(x,y_{obs})f_g(y_{mis}|x,y_{obs}).
\end{equation}

For modeling the mixing probability $\pi_g(x)$ in (\ref{cond}), we adopt the following multinomial logit model:
\begin{equation}\label{pi}
\pi_g(x;u,\alpha)=\frac{u_g\exp(x^T\alpha_g)}{\sum_{h=1}^Gu_h\exp(x^T\alpha_h)}, \ \ \ \ g=1,\ldots,G,
\end{equation}
where $u_g>0$ and $\alpha_g\in \mathcal{R}^q$.
Note that $\log u_g$ can be regarded as an intercept term for the multinomial logistic model.
We set $u_1=1$ and $\alpha_1=0$ for model  identifiability. 
Note that, in model (\ref{pi}), if $u_g\approx 0$, then the $g$th component will be automatically ignored in the model. Hence, by introducing a prior distribution for $u_g$ that has a considerable mass around the origin, we can automatically reduce the number of mixture components in (\ref{cond}).
In view of such sparsity of the model, we refer to the model (\ref{cond}) as sparse conditional Gaussian mixture models.

\section{Bayesian imputation} 

\subsection{Gibbs sampling procedure}

Under the sparse conditional Gaussian mixture model in (\ref{impute}), we wish to develop a Bayesian imputation method by 
specifying the prior distribution of the parameters and describing the computation for the posterior distribution. By choosing suitable prior distributions, the computation for the posterior distribution can be efficient. 
There are three latent components in the Bayesian imputation: the model parameter $\Psi=(u, \alpha, \psi)$ in the joint model of $(z, y)$,  the group-specific latent variable $z$, and the missing data $y_{\rm mis}$ itself. We use Gibb sampling to generate these latent components iteratively.

Regarding the prior for $u_g$ in the multinomial logit model (\ref{pi}), we employ ${\rm Ga}(a, 1)$ for some small $a>0$.
The motivation for the prior distribution of $u_g$ could be presented by considering a sub-model with $\alpha_g=0$. 
When $u_g\sim {\rm Ga}(a, 1)$, the mixing proportion $(u_1/\sum_{h=1}^Gu_h,\ldots,u_G/\sum_{h=1}^Gu_h)$ follows ${\rm Dir}(a,\ldots,a)$, and the Dirichlet distribution has considerable mass on the boundary when $a$ is small positive constant, leading to shrinkage properties for the mixing proportions, as adopted in \cite{linero2018bayesian} and \cite{malsiner2016model} among others.  
Hence, the prior specification for $u_g$ can be regarded as a natural extension of the sparse Dirichlet distribution in the model (\ref{pi}) that depends on some covariate $x$.  
Following \cite{linero2018bayesian} or \cite{malsiner2016model}, we choose $a=1/G$ as a default choice.

For the other parameters, we employ rather standard prior distributions, $B_g\sim {\rm MN}(0, S_{B_1}, S_{B_2})$, $b_g\sim N(0, S_b)$, $\alpha_g\sim N(0, S_{\alpha})$ and $\Sigma\sim {\rm IW}(\nu, S_{\Sigma})$, where $S_{B_1}$ and $S_{\alpha}$ are $q\times q$ matrices, $S_{B_2}$ and $S_{\Sigma}$ are $p\times p$ matrices, $S_b$ and $\nu$ are constants.
Here ${\rm MN}(M, U, V)$ denotes the matrix normal distribution with density proportional to $\exp\{-{\rm tr}[V^{-1}(X-M)^TU^{-1}(X-M)]\}$ and $IW$ denotes the inverse-Wishart distribution.

We let $\Psi$ be the collection of unknown parameters and $\pi(\Psi)$ be the joint prior distribution for $\Psi$. 
Given the dataset $\Dc=\{x_i,y_{i,obs}\}_{i=1,\ldots,n}$, the joint posterior distribution of $\Psi$ and the latent indicator, $z_i$'s, is 
$$
\pi(\Psi,z|\Dc)=\frac{\pi(\Psi)L(z,\Psi|\Dc)}{\int\pi(\Psi)L(z,\Psi|\Dc)d\Psi},
$$
where
$$
L(z,\Psi|\Dc)=\prod_{i=1}^n\Big\{\sum_{h=1}^Gu_h\exp(x_i^T\alpha_h)\Big\}^{-1}
\prod_{g=1}^G\big\{u_g\exp(x_i^T\alpha_g)f_g(y_{i,obs}|x_i; \psi_g\big\}^{I(z_i=g)}.
$$
We employ Markov Chain Monte Carlo methods to generate posterior samples of $z$ and $\Psi$. 
In particular, using Polya-gamma data augmentation \citep{Polson2013}, we can derive an efficient Gibbs sampling algorithm described below. 
\begin{itemize}
\item[-]
{\bf (Sampling of $\alpha_g$)}: \ 
Define $\psi_{ig}=\log u_g + x_i^T\alpha_g - C_{ig}$ with $C_{ig}=\log(\sum_{h\neq g}u_h\exp(x_i^T\alpha_h))$.
Then, the full conditional distribution of $\alpha_g$ and $u_g$ are proportional to 
\begin{equation}\label{PG}
\begin{split}
&\pi(\alpha_g)\pi(u_g)\prod_{i=1}^n\Big\{\sum_{h=1}^Gu_h\exp(x_i^T\alpha_h)\Big\}^{-1}\big\{u_g\exp(x_i^T\alpha_g)\big\}^{I(z_i=g)}\\
&\propto\pi(\alpha_g)\pi(u_g)\prod_{i=1}^n\left\{1+\exp(\psi_{ig})\right\}^{-1}\exp(\psi_{ig})^{I(z_i=g)}\\
& \propto \pi(\alpha_g)\pi(u_g)\prod_{i=1}^n\exp(\psi_{ig}\kappa_{ig})\int_{0}^{\infty}\exp\left(-\frac12\psi_{ig}^2\omega_{ig}\right)PG(\omega_{ig}; 1, 0)d\omega_{ig},
\end{split}
\end{equation}
where $\kappa_{ig}=I(z_i=g)-1/2$, $PG(\cdot; b, c)$ is the density of Polya-gamma distribution with parameters $b$ and $c$ \citep{Polson2013}, and we used the Polya-gamma data augmentation with additional latent variables $\omega_{ig}$ in the final expression. 
The full conditional distribution of $\omega_{ig}$ is $PG(1, \psi_{ig})$, for which we can use the R package ``pgdraw" \citep{pgdraw} to generate the random samples.
Given $\omega_{ig}'s$, the full conditional distribution of $\alpha_g$ is $N(\At_{\alpha_g}\Bt_{\alpha_g}, \At_{\alpha_g})$, where
$$
\At_{\alpha_g}=\left(\sum_{i=1}^n \omega_{ig}x_ix_i^T+S_{\alpha}^{-1}\right)^{-1}, \ \ \ \Bt_{\alpha_g}=\sum_{i=1}^nx_i\big\{\kappa_{ig}+\omega_{ig}(C_{ig}-\log u_g)\big\}.
$$

\item[-]
{\bf Sampling of $u_g$}: \ 
We use the same data augmentation given in (\ref{PG}). 
Although the full conditional distribution is not a familiar form, we can use an efficient independent Metropolis-Hastings algorithm by approximating the prior distribution for $u_g$.
First, we numerically compute $E[\log u_g]$ and ${\rm Var}(\log u_g)$ under the prior distribution $u_g\sim {\rm Ga}(a, 1)$, which are denoted by $\mt$ and $\st^2$, respectively.  
We then approximate the prior distribution of $\log u_g$ as $TN_{(-\infty, 2)}(\mt,\st^2)$, where $TN_{(a,b)}(\mu,\sigma^2)$ is a truncated normal distribution with mean parameter $\mu$ and variance parameter $\sigma^2$ truncated on $(a,b)$. 
Under the approximate prior distribution, the candidate of $\log u_g$ is generated from $TN_{(-\infty,2)}(\At_{u_g}\Bt_{u_g}, \At_{u_g})$, where 
$$
\At_{u_g}=\left(\sum_{i=1}^n\omega_{ig}+\frac1{\st^2}\right)^{-1}, \ \ \ \Bt_{u_g}=\sum_{i=1}^n\kappa_{ig}+\frac{\mt}{\st^2}.
$$
Then, the candidate value $u_g^{\dagger}$ is accepted with probability 
$$
\min\left\{1, \frac{{\rm Ga}(u_g^{\dagger};a,1){\rm TN}_{(-\infty, 2)}(u_g;\mt,\st^2)}{{\rm Ga}(u_g;a,1){\rm TN}_{(-\infty, 2)}(u_g^{\dagger};\mt,\st^2)}\right\}
=
\min\left\{1, \frac{{\rm Ga}(u_g^{\dagger};a,1)\phi(u_g;\mt,\st^2)}{{\rm Ga}(u_g;a,1)\phi(u_g^{\dagger};\mt,\st^2)}\right\}
$$

\item[-]
{\bf Sampling of $b_g$, $B_g$ and $\Sigma_g$}: \ 
The full conditional distribution of $(b_g, B_g)$ is a $p\times (q+1)$ matrix normal distribution, ${\rm MN}(\Mt_g, \Ut_g, \Vt_g)$, where
\begin{align*}
&\ \ \ \ \ 
\Mt_g=\Ut_g^{-1}\left\{\sum_{i=1}^nI(z_i=g)x_{i\ast}y_i^T\right\}\Sigma_g^{-1}\Vt_g^{-1}, \\ 
&\Ut_g=S_{B_2}^{-1} + \Sigma_g^{-1}, \ \ \ \ \ \ 
\Vt_g=S_{B_1}^{-1} + \sum_{i=1}^nI(z_i=g)x_{i\ast}x_{i\ast}^T,
\end{align*} 
and $x_{i\ast}=(1, x_i^T)^T$.
The full conditional distribution of $\Sigma_g$ is ${\rm IW}(\nut_g, \St_{\Sigma_g})$, where 
$$
\nut=\nu +\sum_{i=1}^nI(z_i=g),  \ \ \ \ 
\St_{\Sigma_g}=S_{\Sigma} + \sum_{i=1}^nI(z_i=g)(y_i-b_g-B_gx_i)(y_i-b_g-B_gx_i)^T.
$$

\item[-]
{\bf Sampling of $z_i$}: \ 
The full conditional probability of $z_i=g$ is proportional to $u_g\exp(x_i^T\alpha_g)\phi(y_{i,obs};\beta_g+B_gx_i,\Sigma_g)$.

\item[-]
{\bf Sampling of $y_{i,mis}$}: \ 
Given $z_i=g$, the conditional distribution of $y_{i,mis}$ given $y_{i,obs}$ is the conditional distribution of $y_i\sim N(b_g+B_gx_i, \Sigma_g)$.  
\end{itemize}

\subsection{Imputation under mixed outcomes}\label{sec:mixed}
In practice, it is often the case that $y$ is mixed outcomes, that is, some component is discrete variables such as binary or count.
The proposed method can handle the situation by a slight modification of the model and sampling algorithm.  
Let $y=(y_1,\ldots,y_p)$ is a vector of response variables with mixed scale margins. 
We then introduce a latent continuous variable $y^{\ast}=(y_1^{\ast},\ldots,y_p^{\ast})$ that follows the proposed Gaussian mixture model and express the response $y$ by transforming $y^{\ast}$.
Such approach is often used in modeling discrete variables \citep[e.g.][]{canale2011bayesian,kottas2005nonparametric}.
We first note that we can simply set $y_k=y_k^{\ast}$ when $y_k$ is a continuous scale. 
When $y_k$ is a binary variable, $y_k\in \{0,1\}$ can be induced by $y_k=I(y_k^{\ast}>0)$.
A count variable $y_k\in\{0,1,2\ldots\}$ can be expressed as 
$$
y_k=\sum_{j=0}^{\infty}j I(a_j<y_k^{\ast}\leq a_{j+1}),
$$
where $-\infty=a_0<a_1<a_2<\ldots<\infty$ and a typical choice would be $a_{j+1}=j$.
Other types of variables (e.g. ordered or multinomial variables) can be expressed in a similar manner. 

Under the formulation, the sampling algorithm in Section \ref{sec:SCGM} can still be used given the latent variables, and additional sampling steps for the latent variables should be added. 
For example, under $y_k=I(y_k^{\ast}>0)$, the full conditional distribution of $y_k^{\ast}$ is the conditional distribution of $y_k^{\ast}$ under the model (\ref{cond}) truncated on $(-\infty, 0)$ and $(0,\infty)$ when $y_k$ is $0$ and $1$, respectively. 
Similarly, $y_k^{\ast}$ under the count response can be generated from the conditional distribution of $y_k^{\ast}$ under the model (\ref{cond}) truncated on $(a_j,a_{j+1})$ when $y_k=j$.
For missing responses, the corresponding latent variable $y_k^{\ast}$ can be generated according to the model (\ref{cond}) and transformed to the response scale.

\section{Approximate posterior inference under imputed data}\label{sec:pos-imputed}
Suppose we are interested in estimating a $d$-dimensional parameter $\theta$ which is defined as the minimizer of the loss function $E[L(\theta,Y)]$, where $Y$ is a $p$-dimensional random vector and $y_1,\ldots,y_n$ are independent and identically distributed realizations of $Y$. 
A consistent estimator of $\theta$ is obtained as the minimizer of $\sum_{i=1}^n L(\theta,y_i)$, which is equivalent to the solution of the estimating equation, $\sum_{i=1}^n U(\theta,y_i)$ with $U(\theta,y_i)=\partial L(\theta,y_i)/\partial\theta$, under the assumption that the solution of the estimating equation is unique. 
Since $y$ is subject to missingness, the imputed loss function is 
$$
L(\theta; \Dc)\equiv \int \sum_{i=1}^n L\{\theta,(y_{i,obs},y_{i,mis})\}f(y_{i,mis}|y_{i,obs})dy_{mis}.
$$
where $f(y_{mis}|y_{obs})$ is the predictive posterior distribution of $y_{mis}$ given $y_{obs}$ obtained in the previous section. 
To make uncertainty quantification based on the above objective function, we propose imputed loss-likelihood bootstrap (ILB), described as follows:

\begin{algo}[ILB algorithm]
\label{algo:ILB}
Repeat the following three steps for $B$ times.
\begin{enumerate}
\item
For $i=1,\ldots,n$, generate $y_{i,mis}^{\ast}$ from $f(y_{i,mis}|y_{i,obs})$.

\item
Generate random weight $(w_1,\ldots,w_n)$ independently from the standard exponential distribution ${\rm Exp}(1)$.

\item
Obtain $\theta^{\ast}$ by minimizing $\sum_{i=1}^nw_iL\{\theta, (y_{i,mis}^{\ast},y_{i,obs})\}$.
\end{enumerate}
\end{algo}

Note that the use of the random weight $w_i\sim {\rm Exp}(1)$ is equivalent to using scaled random weight $(w_1/n\bar{w},\ldots,w_n/n\bar{w})$ with $\bar{w}=n^{-1}\sum_{i=1}^n w_i$, which follows ${\rm Dir}(1,\ldots,1)$, since the minimizer of the objective function does not change under rescaling. 
The ILB algorithm can be regarded as an extension of the loss-likelihood bootstrap \citep{lyddon2019general} and weighed likelihood bootstrap \citep{newton1994approximate} to the case with missing observations. 
Note that the ILB algorithm can be easily parallelized, unlike Markov Chain Monte Carlo algorithms.
As shown in \cite{lyddon2019general}, the asymptotic distribution of the loss-likelihood bootstrap is the same as that of general M-estimators. 
For ILB, we can prove the following asymptotic theorem:

\begin{thm}\label{thm:ILB}
Let $\theta^{\ast}$ be an ILB sample obtained from Algorithm \ref{algo:ILB}.
Under regularity conditions, for any Borel set $A\in\mathbb{R}^d$, as $n\to\infty$, we have 
$$
P_{\rm ILB}(\sqrt{n}(\theta^{\ast}-\thh)\in A\mid\Dc_n) \to P(z\in A),
$$
where $\thh={\rm argmin}_\theta L(\theta; \Dc_n)$, and $z\sim N_d(0, J(\theta_0)^{-1}I(\theta_0)J(\theta_0)^{-1})$ with 
\begin{align*}
J(\theta)&=E\left[\int \left(\frac{\partial^2}{\partial \theta\partial\theta^{\top}}L\{\theta, (Y_{obs}, Y_{mis})\}\right) f(Y_{mis}|Y_{obs})dY_{mis}\right], \\
I(\theta)&=E\left[\int \left(\frac{\partial}{\partial \theta}L\{\theta, (Y_{obs}, Y_{mis})\}\right)\left(\frac{\partial}{\partial \theta}L\{\theta, (Y_{obs}, Y_{mis})\}\right)^\top  f(Y_{mis}|Y_{obs})dY_{mis}\right],
\end{align*}
where the expectation is taken with respect to $Y_{obs}$ and $\theta_0={\rm argmin}_{\theta} E[L(\theta, Y)]$. 
\end{thm}

The proof is given in the Appendix. 
Theorem~\ref{thm:ILB} gives an asymptotic justification of ILB, and its asymptotic distribution is the same as $\thh$. 
Hence, based on the ILB samples, $\theta^{(1)},\ldots,\theta^{(B)}$, we can compute the point estimate by the average, $B^{-1}\sum_{b=1}^B \theta^{(b)}$, which is asymptotically equivalent to $\thh$. 
Furthermore, we can readily compute the measure of uncertainty, for example, the posterior standard error or credible interval of $\theta$ based on the ILB samples. 
A notable feature of ILB is that it does not require explicit expressions of $J(\theta)$ and $I(\theta)$, which typically needs complicated algebraic calculations.

When we have a prior distribution $\pi(\cdot)$ for $\theta$, such information can be incorporated into ILB by additionally generating random weight $w_0\sim {\rm Exp}(1)$ and minimizing $\omega\sum_{i=1}^nw_iL\{\theta, (y_{i,mis}^{\ast},y_{i,obs})\}-w_0\log \pi(\theta)$ for some $\omega>0$.
This procedure can be seen as an analogy of weighted Bayesian bootstrap of  \cite{newton2021weighted} for complete data, and similar asymptotic properties can be established.

\section{Simulation studies}\label{sec:sim}

We conduct a simulation study to evaluate the performance of the proposed method together with some existing methods.
We adopted two situations. 
The first one is where all the variables are continuous, while the response variables under the second situation are mixed margins.

\subsection{Continuous outcomes}\label{sec:sim-Gauss}
We adopted the following four scenarios of the data generating process. 

\begin{itemize}
\item[-]
{\bf Scenario 1}: 
We generate $(y_1^{\ast}, y_2^{\ast}, x_1,x_2)$ from a Gaussian mixture model.
For $g\in \{1,2\}$, $P(z_g) = \lambda_g$ with $(\lambda_1,\lambda_2)=(0.4, 0.6)$ and 
$$
(y_1^{\ast}, y_2^{\ast}, x_1,x_2) \mid z=g\sim N(\mu_g, \Sigma),
$$
with $\mu_1= (2,4,1,0)$, $\mu_2=(-2, 7, -3, 0)$, $\Sigma_{(i,j)}= 3\times (-0.5)^{|i-j|}$, where $\Sigma_{(i,j)}$ is the $(i, j)$-element of $\Sigma$.

\item[-]
{\bf Scenario 2}: 
We generate $(x_1,x_2)$ from a Gaussian mixture model with $4$ components, as follows.
For $g=1,\ldots,4$,
\begin{align*}
(x_1,x_2) \mid z=g  \sim N\left(\mu_{g},\left[\begin{array}{cc}
0.5 & 0.1 \\
0.1 & 0.5
\end{array}\right]\right), \ \ \ \ \ 
P(z=g)=\lambda_{g}, 
\end{align*}
where we set $\left(\lambda_{1}, \lambda_{2}, \lambda_{3}, \lambda_{4}\right)=(0.2,0.3,0.2,0.3)$, and $\mu_{1}=(-1,0.5), \mu_{2}=(1,1), \mu_{3}=(0.5,-1)$ and $\mu_4=(0,0)$.
Given $(x_1,x_2)$, we generate $(y_1^{\ast},y_2^{\ast})$ from a conditional Gaussian mixture model with two components, as follows.
\begin{align*}
&y_1^{\ast}=(2+x_1+x_2)I(U_1>c_1) + (-2+0.5x_1-x_2)I(U_1\leq c_1) + \ep_1,\\
&y_2^{\ast}=(10-x_1-x_2)I(U_2>c_2) + (6-0.5x_1+2x_2)I(U_2\leq c_2) + \ep_2,
\end{align*}
where $\ep_1,\ep_2\sim N(0,1)$, $U_1\sim N(1+2x_1+x_2, 1)$, $U_2\sim N(1+x_1+2x_2, 1)$, and $c_k$ is the $60\%$-quantile point of $U_k$ with $k\in\{1,2\}$.

\item[-]
{\bf Scenario 3}:
We use the same generation process for $(x_1,x_2)$, $U_1$ and $U_2$.
Then, $(y_1^{\ast},y_2^{\ast})$ is generated from 
\begin{align*}
&y_1^{\ast}=(2+x_1^2+x_2^2)I(U_1>c_1) + (-2+0.5x_1-x_2^2)I(U_1\leq c_1) + \ep_1,\\
&y_2^{\ast}=(10-x_1^2-x_2)I(U_2>c_2) + (6-0.5x_1+2x_2^2)I(U_2\leq c_2) + \ep_2,
\end{align*}
where $\ep_1,\ep_2\sim N(0,1)$.

\item[-]
{\bf Scenario 4}:
We adopt the same generating model in Scenario 3, except for $\ep_1,\ep_2\sim {\rm Ga}(1,1)$.
\end{itemize}

In this simulation, we set $y_k=y_k^{\ast}$, thereby the response variable $(y_1,y_2)$ are both continuous.
The data generating process of Scenario 1 is a two-component mixture of multivariate normal distributions for four-dimensional variables so that the conditional distribution of $(y_1,y_2)$ given $(x_1, x_2)$ has a form of conditional Gaussian mixtures as assumed in the proposed model.
In Scenarios 3 and 4, the data generating process has sub-group structures that can be approximately modeled by conditional Gaussian mixtures. 
However, in scenario 4, the conditional distributions of $(y_1,y_2)$ is not Gaussian.

We generate 500 finite population data with the population size $N=10,000$ and select a sample of size $n$ equal to 1000 using simple random sampling from the finite population. 
Once the full sample is obtained, we generate the response indicator $(\delta_{1i},\delta_{2i})$ from ${\rm logit}\{P(\delta_{1i}=1)\} =  1.5-0.5x_{1i}$ and ${\rm logit}\{P(\delta_{2i}=1)\} =  1-0.5x_{2i}$. 
We assume that $y_{ki}$ is observed only when $\delta_{ki}=1$. 
The overall missing rate is about 30\%.

For each realized incomplete sample, we applied the following methods to impute the missing values.

\begin{itemize}
\item[-]
{\bf PMM}: Predictive-mean matching commonly used for multiple imputation using the chained equations process. 

\item[-]
{\bf LR}: An iterative method imputing missing values using linear regression.

\item[-]
{\bf RF}: An iterative method imputing missing values using random forest.

\item[-]
{\bf SCGM}: The semiparametric imputation using conditional Gaussian mixture \citep{lee2022semiparametric}, where the number of components $G$ is selected from $\{1,\ldots,7\}$ by the BIC.

\item[-]
{\bf BLGM}: The proposed Bayesian latent sparse Gaussian mixture with $7$ components. 
We generated 1500 posterior samples of the missing values after discarding the first 500 samples as burn-in samples.
\end{itemize}

For the first three methods, we used R package \verb+mice+ \citep{buuren2011mice} with default settings. 
To evaluate the imputation accuracy, we computed the mean absolute error
(MAE) defined as follows:  
\begin{align*}
{\rm MAE}_k=\bigg\{\sum_{i=1}^{n}(1-\delta_{ki})\bigg\}^{-1}\sum_{i=1}^n(1-\delta_{ki})|\yh_{ki}^{\ast}-y_{ki}|,  \ \ \ k\in \{1,2\}.
\end{align*}
where $\yh_i^{\ast}$ is the imputed value of missing $y_i$ with $\delta_i=0$ and $y_i$ is the true value.
We also computed estimates of the finite population means, $\theta_k$, that can be estimated as 
$$
\widehat{\theta}_k=\frac1n\sum_{i=1}^n\Big\{\delta_{ki}y_{ki}+(1-\delta_{ki})\yh_{ki}^{\ast}\Big\}, \ \ \ k\in \{1,2\}.
$$

\begin{figure}[t]
\centering
\includegraphics[width=13cm,clip]{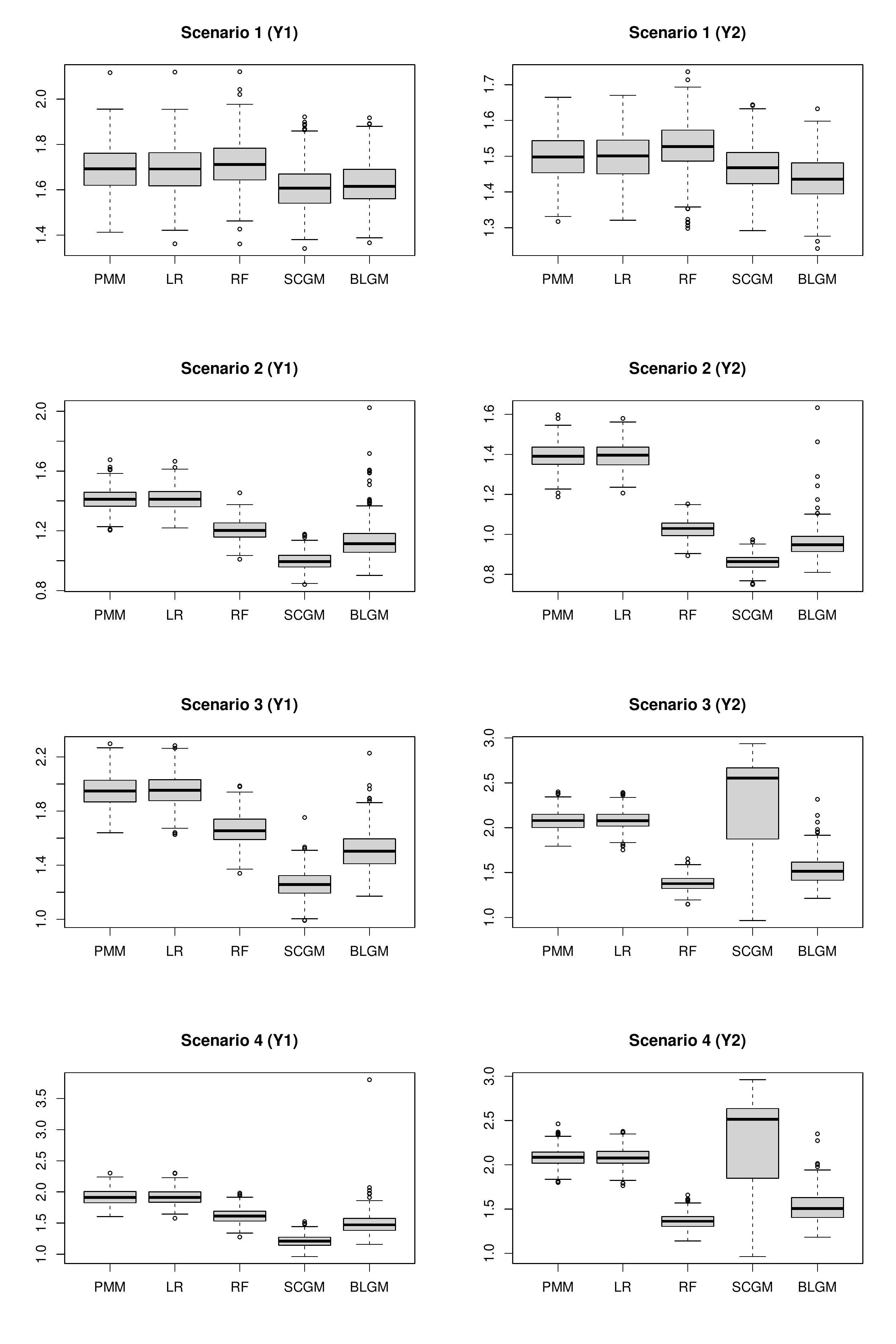}
\caption{Boxplots of MAE values of the imputation based on five methods when the response variables are both continuous-valued.  
\label{fig:sim-Gauss}
}
\end{figure}

\begin{figure}[t]
\centering
\includegraphics[width=13cm,clip]{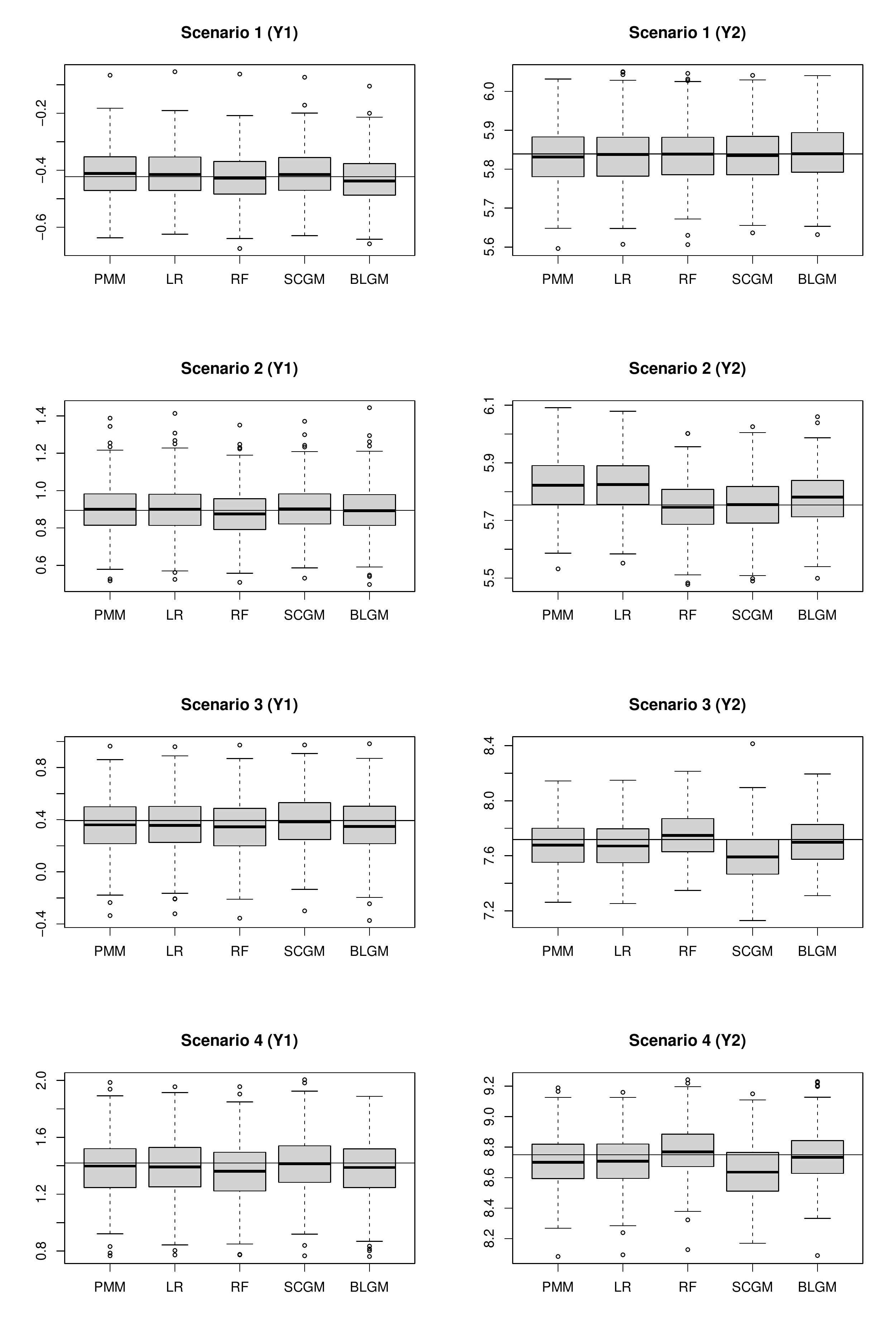}
\caption{Boxplots of estimates of the population means based on the five methods when the response variables are both continuous-valued.    
\label{fig:sim-Gauss-est}
}
\end{figure}

The boxplots of MAE values for the five models are given in Figure \ref{fig:sim-Gauss}.
It shows that the proposed BLGM method produces more accurate imputation than the MICE-based methods.
On the other hand, the two methods, SCGM and BLGM, seem comparable in that SCGM performs slightly better than BLGM in some cases but considerably worse in the imputation of $y_2$ in Scenarios 3 and 4.  
This is not surprising because both SCGM and BLGM are based on conditional Gaussian mixtures.   
The boxplots of the estimated population mean under 500 replications are presented in Figure \ref{fig:sim-Gauss-est}, which shows that SCGM and BCGM are quite comparable, and they tend to provide more accurate estimates than the other MICE-based methods. 
In Table \ref{tab:selection}, we reported the average number of the non-null components in the BLGM method, where a component in which at least one sample is included is regarded as non-null. 
The result reveals that the BLGM method can adaptively reduce the number of non-null components owing to the sparsity-inducing formulation of $u_g$, depending on the complexity of the underlying data generating structure.

We also considered inference on the population mean $\theta$.
Using the quadratic loss function, $L(\theta_k, y_k)=\sum_{i=1}^n(y_{ki}-\theta_k)^2 (k=1,2)$, and the ILB algorithm described in Section \ref{sec:pos-imputed}, we obtained approximate posterior samples of $\theta$ and then computed 95\% credible intervals of $\theta_k$. 
The empirical coverage probabilities based on 500 replications are shown in the left column of Table \ref{tab:CP}.
The coverage probability is around the nominal level, indicating that the proposed method can produce valid statistical inference on the parameter of interest by flexibly imputing the missing values.

\begin{table}[htbp]
\caption{Coverage probabilities of $95\%$ credible intervals of population means based on the ILB algorithm.}
\label{tab:CP}
\begin{center}
\begin{tabular}{ccccccccccc}
\hline
& &\multicolumn{2}{c}{Continuous case} & &\multicolumn{2}{c}{ Mixed case}\\
 &  & $y_1$ & $y_2$ &  & $y_1$(binary) & $y_2$ (count) \\
\hline
Scenario 1 &  & 96.2 & 96.8 &  & 94.0 & 94.8 \\
Scenario 2 &  & 95.0 & 95.6 &  & 94.6 & 94.4 \\
Scenario 3 &  & 94.8 & 96.0 &  & 96.6 & 92.4 \\
Scenario 4 &  & 95.4 & 95.2 &  & 96.2 & 93.8 \\
\hline
\end{tabular}
\end{center}
\end{table}

\begin{table}[htbp]
\caption{The average number of the non-null components in the BLGM based on 500 replications, where the number of components is 7.}
\label{tab:selection}
\begin{center}
\begin{tabular}{ccccccccccc}
\hline
Outcome & Scenario 1 & Scenario 2 & Scenario 3 & Scenario 4 \\
\hline
Continuous & 1.90 & 3.36 & 4.05 & 4.21 \\
Mixed & 2.07 & 3.24 & 3.80 & 3.77 \\
\hline
\end{tabular}
\end{center}
\end{table}

\subsection{Mixed outcomes}
We next consider a mixed outcome situation.
Using the same missing mechanism and data generating process for the latent variables $y_k^{\ast}$ as in Section \ref{sec:sim-Gauss}, we define the outcome variables as $y_1=I(y_1^{\ast})$ and $y_2=\sum_{j=0}^{\infty}jI(a_j<y_2^{\ast}\leq a_{j+1})$, where $a_0=-\infty$ and $a_{j+1}=j$.
Hence, the two response variables $y_1$ and $y_2$ are binary and count, respectively. 
To impute the missing values, we employed the three MICE-based methods (PMM, LR, and RF), where logistic and Poisson regression models are used when applying LR for binary and count outcomes, respectively.
Note that the SCGM cannot handle non-continuous responses so that we omitted it from our comparison. 
In applying the BLGM method, we set 7 components and generated 1500 posterior samples of the missing values after discarding the first 500 samples.
Based on the imputed values, we also computed the estimates of the population means.
The performance of the imputation is evaluated by the mean classification error (MCE) given by 
\begin{align*}
{\rm MCE}=\bigg\{\sum_{i=1}^{n}(1-\delta_{1i})\bigg\}^{-1}\sum_{i=1}^n(1-\delta_{1i})I(\yh_{1i}^{\ast}\neq y_{1i}), 
\end{align*}
for binary responses, and a modified version of mean absolute percentage error (MAPE) given by 
\begin{align*}
{\rm MAPE}=\bigg\{\sum_{i=1}^{n}(1-\delta_{2i})\bigg\}^{-1}\sum_{i=1}^n(1-\delta_{2i})\frac{|\yh_{2i}^{\ast}-y_{2i}|}{1+y_{2i}}, 
\end{align*}
for count responses.

\begin{figure}[t]
\centering
\includegraphics[width=13cm,clip]{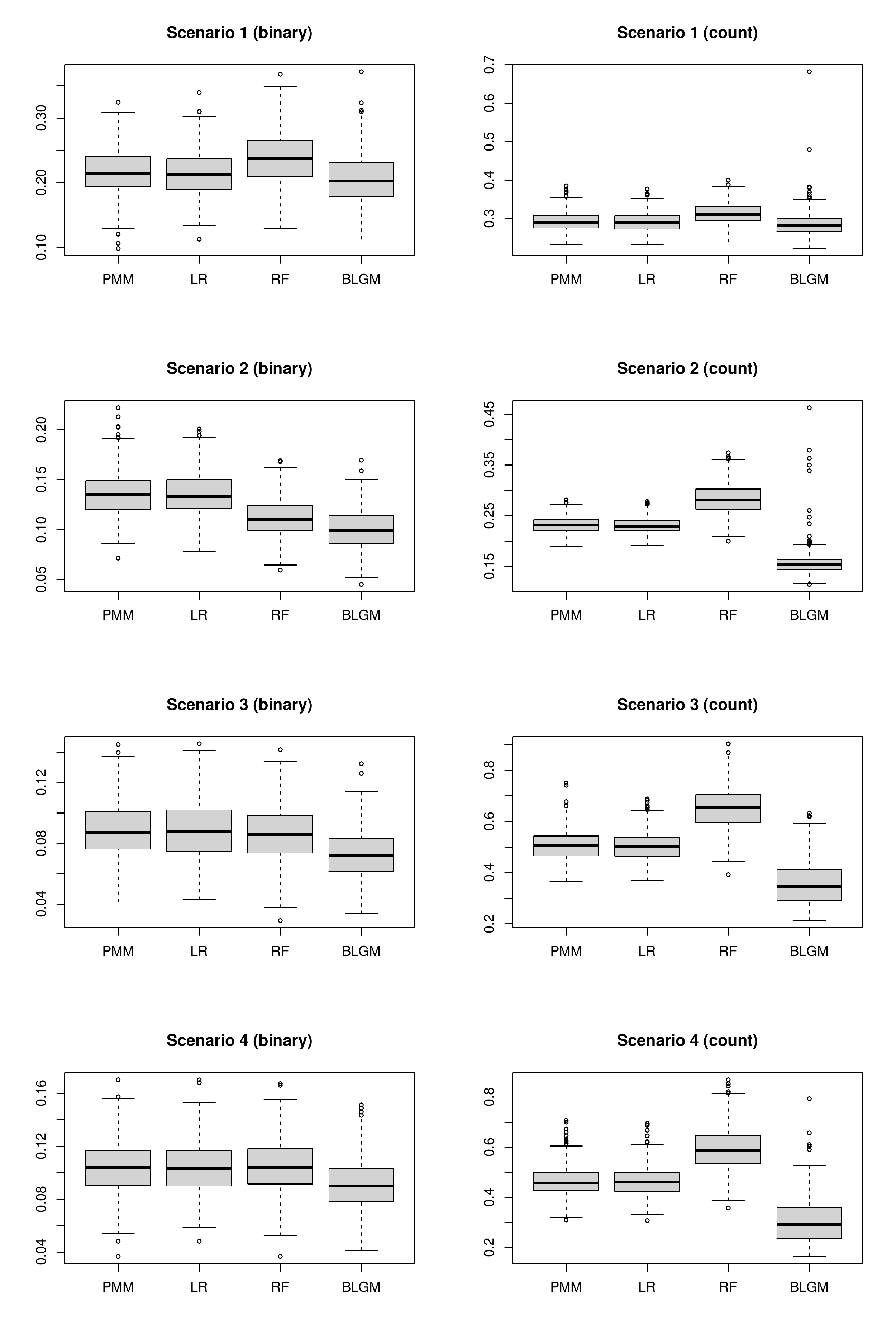}
\caption{Boxplots of MCE (for binary responses) and MAPE (for count responses) values of the imputation based on the four methods when the response variables are mixed margins. 
\label{fig:sim-Mix}
}
\end{figure}

\begin{figure}[t]
\centering
\includegraphics[width=13cm,clip]{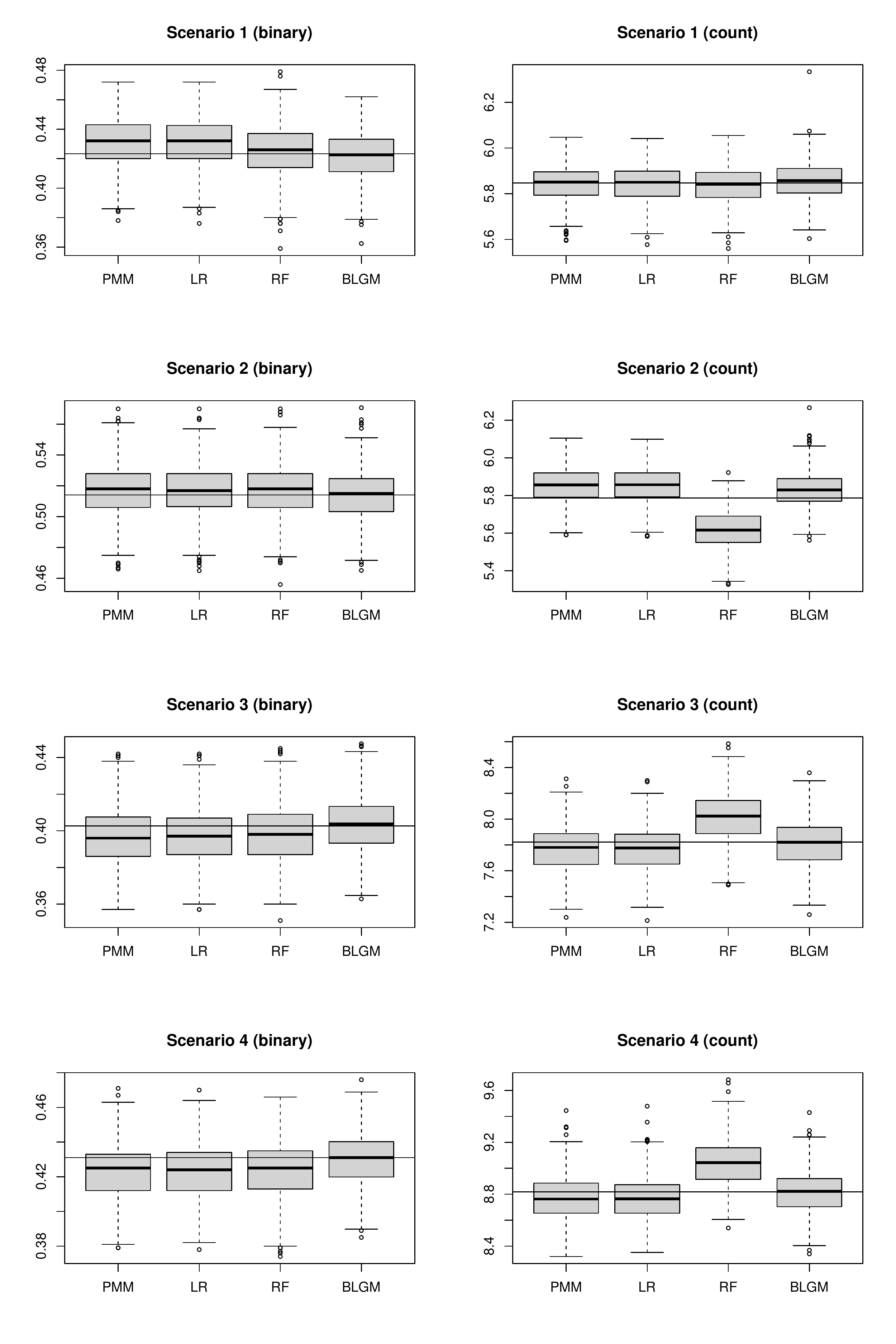}
\caption{Boxplots of estimates of the population means based on the five methods when the response variables are mixed margins.
\label{fig:sim-Mix-est}
}
\end{figure}

The boxplots of MCE and MAPE are reported in Figure \ref{fig:sim-Mix}.
It is observed that the overall performance of the proposed BLGM method is superior to that of the others. 
In particular, the BLGM method is appealing when the true data generating process is complex, as in Scenarios 2-4.  
We also present the average number of the non-null components in Table \ref{tab:selection}, which shows the adaptive property of the BLGM method as confirmed in the continuous case simulation in Section \ref{sec:sim-Gauss}.  
Figure \ref{fig:sim-Mix-est} presents the estimated population means of the four methods. 
It can be seen that BLGM provides almost unbiased estimates in all the scenarios while the other methods produce slightly biased estimates. 
Especially, the RF method produces considerably biased results in Scenarios 2-4. 
Using the same quadratic loss function for the population mean $\theta_k$ of $y_k$ and the ILB algorithm, we obtained $95\%$ credible intervals of $\theta_k$.
The empirical coverage probabilities based on $500$ replications are presented in the right row of Table~\ref{tab:CP}, which shows that valid inference on $\theta_k$ can be done under mixed outcomes.

\section{Illustration using real data}
We illustrate the proposed method using the 2017 Korean Household Income and Expenditure Survey (KHIES) conducted by Statistics Korea.
One purpose of the KHIES is to provide the latest information about Korean household welfare-related status, which measures several different types of income items per person in a household, expenditure-related items, and basic demographic information. 
The sample size of the dataset is $n=17711$.
A more detailed description of the dataset can be found in \cite{lee2022semiparametric}.

In our illustration, we use four variables, administrative income, survey income, age, and education, where education is an order variable taking values from $1$ to $7$.
Here, administrative income is the primary study variable in our illustration and is subject to missingness. 
The overall missing rate is $15\%$. 
Since the observed administrative income and survey income are similar values, survey income will be a useful covariate for administrative income of interest.
To illustrate the effectiveness of imputation under mixed outcomes, we randomly omitted the observed values of education and made statistical inference on the mean and quantiles of administrative income.
Let $y_1$ and $y_2$ be education and administrative income (scaled by 1000), respectively, and let $x_1$ and $x_2$ be age and survey income (scaled by 1000), respectively. 
To introduce missingness in $y_1$, we consider the following three scenarios of missing probability: 
\begin{align*}
&\text{(Scenario 1)} \ \ \ \ 
{\rm logistic}(-8+0.01x_1+x_2), \ \ \ \ 
\text{(Scenario 2)} \ \ \ \ 
{\rm logistic}(-6-0.01x_1+x_2),\\
&\text{(Scenario 3)} \ \ \ \ 
{\rm logistic}\{-6-0.1x_1+(x_2-\mu)^2\},
\end{align*}
where $\mu$ is the sample mean of $x_2$. 
Note that overall missing rates in the above three scenarios are around $10\%$.
Here we set our goals to making statistical inference on marginal $25\%, 50\%$ and $75\%$ quantiles of $y_2$ and quadratic regression for $y_2$ given $y_1$.

To impute the missing observations in $(y_1,y_2)$, we applied the proposed BLGM method of the form: 
$$
f(y_1,y_2|x_1,x_2)=\sum_{g=1}^G \pi_g(x_1,x_2)
\left(\begin{array}{c}
h(\beta_{g10}+\beta_{g11}x_1+\beta_{g12}x_2+\ep_{g1})\\
\beta_{g20}+\beta_{g21}x_1+\beta_{g22}x_2+\ep_{g2}
\end{array}\right), 
$$
where $(\ep_{g1}, \ep_{g2})\sim N((0,0), {\rm diag}(\sigma_{g1}^2,\sigma_{g2}^2))$ and $h(u)=\sum_{j=1}^7 jI(a_{j-1}<u\leq a_{j})$ with $a_{j}=j$ for $j=1,\ldots,6$, $a_0=-\infty$ and $a_7=\infty$.
We set $G=8$ and use the model (\ref{pi}) for $\pi_g(x_1,x_2)$.
Based on 8000 posterior samples after discarding the first 2000 samples, we computed posterior means of mixing proportions of $G=8$ groups, defined as 
$$
\Pi_g=\frac{1}{n}\sum_{i=1}^nI(z_i=g), \ \ \ g=1,\ldots,G,
$$
where $z_i$ is the latent grouping variable whose conditional distribution is given in (\ref{cond.prob}).
Furthermore, posterior means of $\beta_{g22}$, a regression coefficient of $y_2$ (administrative income) on $x_2$ (survey income), and $\sigma_{g2}^2$, an error variance of $y_2$. 
The results are presented in Table~\ref{tab:app1}.
It is observed that the number of non-empty components is $4$ (Scenario 1) and $5$ (Scenario 2 and 3), and the remaining components are empty owing to the sparsity-inducing prior for $u_g$ controlling the overall contribution of group-wise models as introduced in (\ref{pi}).   
This result suggests that the proposed method can automatically identify the necessary number of components by specifying a sufficiently large number of components, $G$.  
We can also see that there is a component in which $\beta_{g22}$ and $\sigma_{g2}$ are estimated as 1.00 and 0.00, respectively, under all the scenarios (as presented in bold in Table~\ref{tab:app1}), suggesting that the mixture model successfully distinguishes subsamples having the same values of survey and administrative incomes ($x_2$ and $y_2$), from the other samples. 
Furthermore, a small part of the samples have much larger administrative income than survey income and have a large variance (as presented in italic in Table~\ref{tab:app1}).

For comparison, we employed two MICE-based imputation methods, PMM and RF, as used in Section~\ref{sec:sim}. 
Note that SCGM cannot be used here because $(y_1, y_2)$ has mixed outcomes.  
The performance of imputation is measured by the mean squared errors (MSE) of the imputed observations, defined as 
$$
{\rm MSE}=\bigg\{\sum_{i=1}^n (1-\delta_{1i})\bigg\}^{-1}\sum_{i=1}^n(1-\delta_{1i})(y_{1i}^{\ast}-y_{1i})^2,
$$
where $\delta_{1i}$ is an response indicator for $y_{1i}$ and $y_{1i}^{\ast}$ is the imputed value of $y_{1i}$.
The results under three scenarios are given in Table~\ref{tab:app2}, which shows the superior performance of BLGM to PMM and RF.

\begin{figure}[!htb]
\centering
\includegraphics[width=16cm,clip]{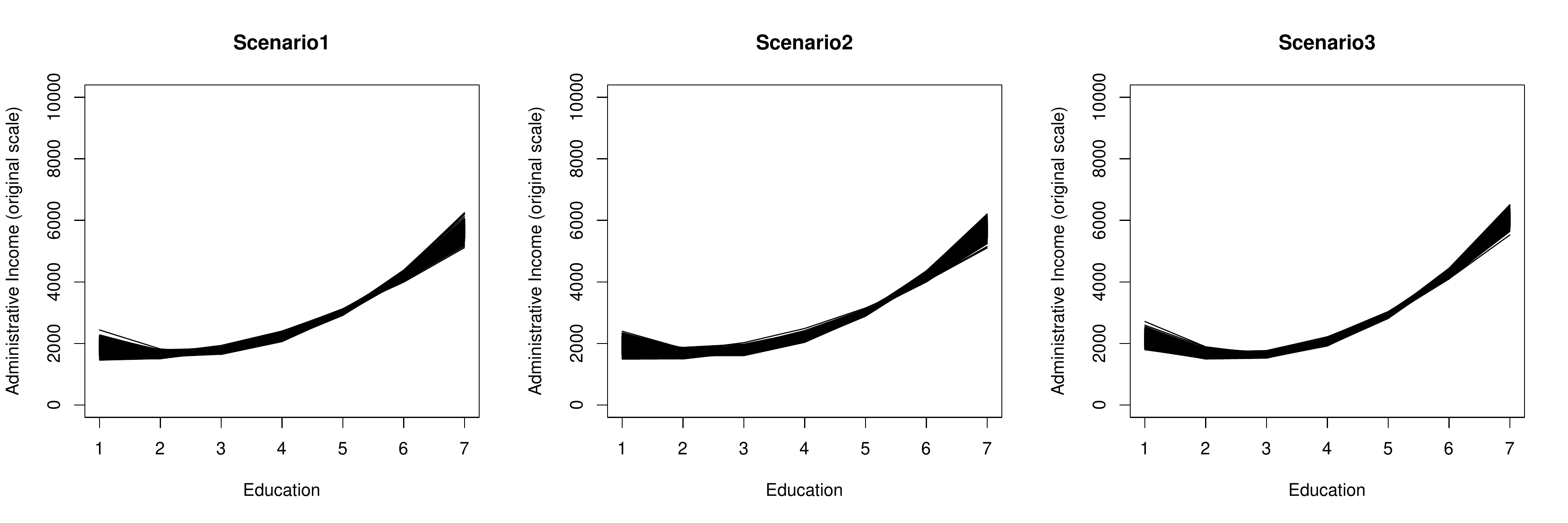}
\caption{Posterior samples of quadratic regression curves for administrative income as a function of the education level. 
\label{fig:app-reg}
}
\end{figure}

We next carry out statistical inference on parameters defined by the distributions of $(y_1 ,y_2)$. 
We first consider marginal quantiles of administrative income ($y_2$), defined as the minimizer of the check loss function, $L_{\tau}(\theta, y)=\sum_{i=1}^n (y-\theta)(\tau-I(y-\theta<0))$.
Using the ILB algorithm in Section~\ref{sec:pos-imputed}, we obtained posterior samples of $\theta$ for $\tau=0.25, 0.5, 0.75$, corresponding to $25\%, 50\%$ and $75\%$ quantiles of $y_2$. 
In Table~\ref{tab:app2}, we reported posterior means and $95\%$ credible intervals of the three quantiles under three scenarios. 
For comparison, we also present estimates based on survey income ($x_2$) and imputed administrative income obtained by PMM and RF. 
First, it is observed that BLGM provides almost identical results under all the scenarios, indicating the flexibility of BLGM as an imputation model.
Secondly, PMM and RF tend to produce larger (smaller) estimates of $25\%$ ($75\%$) quantile level than BLGM, implying that PMM and RF may underestimate the variability of $y_2$ compared with BLGM.
Finally, we estimate a quadratic regression model by minimizing $L(\theta, y)=\sum_{i=1}^n (y_{2i}-\theta_0-\theta_1y_{1i}-\theta_2y_{1i})^2$.
We again use the ILB algorithm to generate posterior samples of $\theta$. 
The quadratic regression curves for each sampled $\theta$ are given in Figure~\ref{fig:app-reg}, showing that the results are almost identical over three missing scenarios and the posterior samples of the quadratic curves seem to exhibit reasonable amount of uncertainty.

\begin{table}[htbp]
\caption{Posterior means of mixing proportion ($\Pi_g$), regression coefficient of $y_2$ on $x_2$ ($\beta_{g22}$), and error variance of $y_2$ ($\sigma_{g2}^2$). 
}
\label{tab:app1}
\begin{center}
\begin{tabular}{cccccccccccccc}
\hline
&& \multicolumn{3}{c}{Scenario 1} && \multicolumn{3}{c}{Scenario 2} && \multicolumn{3}{c}{Scenario 3}\\
Group& & $\Pi_g$ & $\beta_{g22}$ & $\sigma_{g2}^2$ &  & $\Pi_g$ & $\beta_{g22}$ & $\sigma_{g2}^2$ &  & $\Pi_g$ & $\beta_{g22}$ & $\sigma_{g2}^2$ \\
 \hline
1 &  & 0.45 & 0.98 & 0.74 &  & 0.54 & 1.04 & 1.14 &  & 0.28 & 0.91 & 1.77 \\
2 &  & {\bf 0.26} & {\bf 1.00} & {\bf 0.00} &  & {\bf 0.26} & {\bf 1.00} & {\bf 0.00} &  & 0.26 & 1.06 & 1.53 \\
3 &  & 0.24 & 0.23 & 0.57 &  & 0.11 & 1.01 & 0.10 &  & {\bf 0.24} & {\bf 1.00} & {\bf 0.00} \\
4 &  & {\it 0.04} & {\it 1.26} & {\it 43.28} &  & 0.05 & 0.22 & 0.90 &  & 0.19 & 0.97 & 0.09 \\
5 &  & 0.00 & -- & -- &  & {\it 0.04} & {\it 1.36} & {\it 45.25} &  & {\it 0.03} & {\it 1.37} & {\it 68.11} \\
$6\sim 8$ &  & 0.00 & -- & -- &  & 0.00 & -- & -- &  & 0.00 & -- & -- \\
\hline
\end{tabular}
\end{center}
\end{table}

\begin{table}[htbp]
\caption{Mean squared errors (MSE) of imputed items for $y_1$ (education).}
\label{tab:app2}
\begin{center}
\begin{tabular}{ccccccccccc}
\hline
&& Scenario 1 & Scenario 2 & Scenario 3\\
\hline
PMM &  & 1.418 & 1.346 & 1.283 \\
RF &  & 1.477 & 1.373 & 1.443 \\
 BLGM &  &  1.248 &  1.174 &  1.191 \\
\hline
\end{tabular}
\end{center}
\end{table}

\begin{table}[htbp]
\caption{Imputation results with $95\%$ confidence interval and estimates of survey incomes in the original scale (Unit: KRW 1,000).
}
\label{tab:app3}
\begin{center}
\begin{tabular}{ccccccccccc}
\hline
& Quantile & Survey  & PMM &RF & \multicolumn{2}{c}{BLGM} \\
& level & estimate& estimate& estimate & estimate & $95\%$ interval\\
\hline
&$25\%$  & 1445 & 1230 & 1300 & 1203 & (1200, 1224) \\
Scenario 1 & $50\%$  & 2400 & 2230 & 2287 & 2232 & (2186, 2274) \\
&$75\%$  & 4000 & 4139 & 4111 & 4167 & (4092, 4243) \\
\hline
& $25\%$  & 1445 & 1229 & 1300 & 1202 & (1200, 1221) \\
Scenario 2 & $50\%$ & 2400 & 2241 & 2306 & 2249 & (2203, 2290) \\
&$75\%$ & 4000 & 4150 & 4147 & 4165 & (4083, 4243) \\
\hline
&$25\%$ & 1445 & 1227 & 1292 & 1202 & (1200, 1224) \\
Scenario 3 &$50\%$ & 2400 & 2238 & 2285 & 2254 & (2206, 2295) \\
&$75\%$ & 4000 & 4145 & 4111 & 4173 & (4090, 4252) \\
\hline
\end{tabular}
\end{center}
\end{table}

\section{Discussion}
When the dimension of $x$ is large, we may employ techniques of Bayesian sparse estimation using shrinkage priors such as horseshoe prior \citep{carvalho2009handling} for each element of the coefficient matrix $B_g$.
Since the full conditional distribution of $B_g$ is matrix normal, as noted in Section \ref{sec:SCGM}, we can still develop an efficient Gibbs sampling algorithm using the mixture representation of the horseshoe prior.
Furthermore, when the missing mechanism is nonignorable, we may still use the BLGM combined with the flexible Bayesian selection model \citep[e.g.][]{sugasawa2022bayesian}, which might be valuable for future research.

\section*{Acknowledgement}
This work is partially supported by the Japan Society for the Promotion of Science (JSPS KAKENHI) grant number 21H00069 and the MEXT Project for Seismology toward Research Innovation with Data of Earthquake (JPJ010217).

\appendix
\vspace{0.5cm}
\begin{center}
{\bf \large Appendix }
\end{center}
\section*{Proof of Theorem~\ref{thm:ILB}}

We assume the regularity conditions given in the proof of Theorem~1 of \cite{lyddon2019general}, for the imputed loss function $\int L(\theta, y)f(y_{mis}|y_{obs})dy_{mis}$. 
In what follows, we skip detailed evaluation of the reminder term of asymptotic expansions, where their evaluation can be done in the same way as done in the proof of Theorem~1 of \cite{lyddon2019general}.

Given the random weight $w$ and missing observations $y_{i,mis}$, the estimating equation for $\theta$ is 
$$
\sum_{i=1}^n w_i U\{\theta,(y_{i,obs}, y_{i,mis})\}=0,
$$
where $U\{\theta,(y_{i,obs}, y_{i,mis})\}=\partial L\{\theta,(y_{i,obs}, y_{i,mis})\}/\partial\theta$.
Expanding the estimating function around $\theta=\thh$, we have 
\begin{align*}
\sum_{i=1}^n w_i & U\{\theta,(y_{i,obs}, y_{i,mis})\}\\
&=\sum_{i=1}^n w_i U\{\thh,(y_{i,obs}, y_{i,mis})\} + \sum_{i=1}^n w_i U'\{\thh,(y_{i,obs}, y_{i,mis})\}(\theta-\thh) +  R_n
\end{align*}
where $U'\{\th,(y_{i,obs}, y_{i,mis})\}=\partial U'\{\th,(y_{i,obs}, y_{i,mis})\}/\partial\theta^{\top}$ and $R_n$ is a reminder term. 
Since the left-hand side is $0$ at $\theta=\theta^{\ast}$, we have 
\begin{align*}
\theta^{\ast}-\thh=\left[\sum_{i=1}^n w_i U'\{\thh,(y_{i,obs}, y_{i,mis})\}\right]^{-1}\sum_{i=1}^n w_i U\{\thh,(y_{i,obs}, y_{i,mis})\} + R'_n,
\end{align*}
where $R_n'$ is another reminder term. 
Note that $y_{1,mis},\ldots,y_{n,mis}$ are independent given $\Dc_n$. 
Since $w_1,\ldots,w_n$ are independent and $E[w_i]=1$, it follows that 
\begin{align*}
\frac1n\sum_{i=1}^n& w_i U'\{\thh,(y_{i,obs}, y_{i,mis})\}\\
&\to \lim_{n\to\infty}\left[ \frac1n\sum_{i=1}^n \int U'\{\thh,(y_{i,obs}, y_{i,mis})\}f(y_{i,mis}|y_{i,obs})dy_{i,mis}\right]=J(\theta_0),
\end{align*}
as $n\to\infty$ under given $\Dc_n$.
Furthermore, as $n\to\infty$, we have
\begin{align*}
\frac1n\sum_{i=1}^n& w_i U\{\thh,(y_{i,obs}, y_{i,mis})\}\\
&\to \lim_{n\to\infty}\left[ \frac1n\sum_{i=1}^n \int U\{\thh,(y_{i,obs}, y_{i,mis})\}f(y_{i,mis}|y_{i,obs})dy_{i,mis}\right]=0,
\end{align*}
and 
\begin{align*}
&\frac1n\sum_{i=1}^n w_i^2 U\{\thh,(y_{i,obs}, y_{i,mis})\}U\{\thh,(y_{i,obs}, y_{i,mis})\}^\top\\
& \to \lim_{n\to\infty}\left[ \frac1n\sum_{i=1}^n \int U\{\thh,(y_{i,obs}, y_{i,mis})\}U\{\thh,(y_{i,obs}, y_{i,mis})\}^\top f(y_{i,mis}|y_{i,obs})dy_{i,mis}\right]=I(\theta_0),
\end{align*}
since $E[w_i^2]=1$.
Then, $n^{-1/2}\sum_{i=1}^n w_i U\{\thh,(y_{i,obs}, y_{i,mis})\}$ is asymptotically normal with mean $0$ and variance-covariance matrix $I(\theta_0)$, which completes the proof.

\vspace{1cm}
\bibliographystyle{chicago}
\bibliography{ref}

\end{document}